# A Spark ML – driven preprocessing approach for deep learning-based scholarly data applications

Samiya Khan[a,1], Xiufeng Liu[b], Mansaf Alam[a,2]

[a]*Jamia Millia Islamia, New Delhi, India*
[b]*Technical University of Denmark, Denmark*

**Highlights**

- Proposes a Spark ML-driven approach for preprocessing textual data, which is used as input for training application-specific deep learning models.
- Presents implementation of four APIs for Spark ML feature class. These APIs perform common text cleaning tasks and can be used as stages in preprocessing pipeline.
- Specifies details related to implementation of 'title or summary generation from abstracts using deep learning', which is chosen as the case study for getting experimental results.
- Provides a testing and validation study, providing investigations related to ingestion time, preprocessing time, cumulative time, accuracy and cost-benefit.

**Abstract:** Big data has found applications in multiple domains. One of the largest sources of textual big data is scientific documents and papers. Big scholarly data have been used in numerous ways to create innovative applications such as collaborator discovery, expert finding and research management systems. With the advent of advanced machine and deep learning techniques, the accuracy and novelty of such applications have risen manifold. However, the biggest challenge in the development of deep learning models for scholarly applications in cloud-based environment is the underutilization of resources because of the excessive time taken by textual preprocessing. This paper presents a preprocessing pipeline that makes use of Spark for data ingestion and Spark ML for pipelining preprocessing tasks. The evaluation of the proposed work is done using a case study, which uses LSTM-based text summarization for generating title or summary from abstract of any research. The ingestion, preprocessing and cumulative time for the proposed approach (P3SAPP) is much lower than the conventional approach (CA), which manifests in reduction of costs as well.

---

[1]

1. **Introduction**

With a growing demand for automation in this age of artificial intelligence, many application domains start to adopt this technology [50]. Education and research can use big data-enabled solutions in diverse applications [51][52]. Scholarly applications are one such a domain that makes extensive use of natural language processing at the backend with machine learning and deep learning to develop innovative applications [1] for researchers. From recommender systems [2] to text summarization mechanisms [3], implementation of complex analytical techniques on big scholarly data can be used to solve different purposes for the common benefit of research community.

A classic example of this assertion is automatic keyword extraction [4], which extracts keywords from scholarly articles using multiple parameters, such as the frequency of key term occurrence [5], citations [6], author relationships, and more. In addition, research paper recommenders [7], venue recommendation systems [8] and analytically - enabled research management systems [9] are some of the other existing applications of big scholarly data analytics.

Scholarly big data comprise textual and image data. Different applications are expected to use big scholarly data in different ways. For instance, if a research paper writing system offers image caption generation as a service, then the images of a scholarly article shall be used. In this regard, a majority of scholarly applications focus on textual data analytics, for which the required text is extracted from a PDF or HTML webpage, cleaned using NLP (Natural Language Processing) algorithms [10] and used as the input to create diverse machine learning and deep learning-based applications.

The problem with the development of deep learning models for natural language processing-based scholarly applications is that the preprocessing stage is extremely resource-intensive, and time consuming. In a cloud-based development environment, the GPU purchased for the project remains underutilized during this time. The increase in time for an underutilized resource increases the project cost.

The motivation of this work is to facilitate the development of novel scholarly applications, in which natural language processing and deep learning have been used. This paper proposes a framework based on a preprocessing pipeline. This framework uses Spark ML for implementing APIs and parallelizing the different stages of data preparation in scholarly applications, which can greatly improve programmer productivity and reduce project cost. The running of the proposed framework also has a high efficiency, in exchange for a minimal loss of accuracy.        .

The rest of the paper is organized as follows: Section 2 introduces the background of scholarly data applications and conventional implementation patterns. Section 3 describes the proposed approach, P3SAPP, also explaining the need for this research. Section 4 elucidates the methodology used for implementation of the proposed preprocessing pipeline. Section 5 presents the results of the experiments performed along with a cost benefit analysis of the proposed approach (P3SAPP) and its comparison with conventional approach (CA).  Section 6 presents the conclusions and scope for future research in this area.

2. **Background**

Big scholarly data consists of documents, typically in PDF format or HTML format, with a structure having several sections, including abstract, keywords, body and references. The constituents of these structures are essentially unstructured with text, images and tables present. Different structures make use of data from different or all sections. For instance, in order to create a citation graph [11], the text present in the references section needs to be scanned. In addition to the above, scholarly data is also generated as a result of user-system interaction in the form of logs. This data can be investigated for demographic analysis, system statistics and markers related to users and usage. Pig and Hive can be used for such analytics [12].

With that said, a major portion of big scholarly data comes from scholarly documents. Khan et al. [13] divides big scholarly data applications, on the basis of functionality, into five categories namely collaborator discovery, research management, expert finding systems, user logs analysis and other recommender systems. All these applications require data to be extracted from PDFs or webpages and bifurcated into sections; out of which relevant sections for the analysis concerned are chosen and acted upon. For example, collaborator discovery and expert finding mainly depend on author information and citation text. The rest of the textual information available in the paper holds less value for this case.

The big data lifecycle for scholarly applications is illustrated in Fig. 1. The first step is the acquisition that deals with acquiring scholarly documents in the form of PDFs or webpages [14]. The data from these sources are extracted and collated into JSON or XML files. This process is referred to as 'extraction' [15]. Specific applications require specific textual information. Moreover, the variations in format and data provided in a research article cause repeated and variable occurrence of nulls in the extracted data. Besides, multiple copies and versions of scholarly articles are available on the Internet. This makes it highly likely that duplicates are present. Regardless of the application, these preprocessing steps are essential for working with big scholarly data.

Fig. 1. Big Scholarly Data Lifecycle

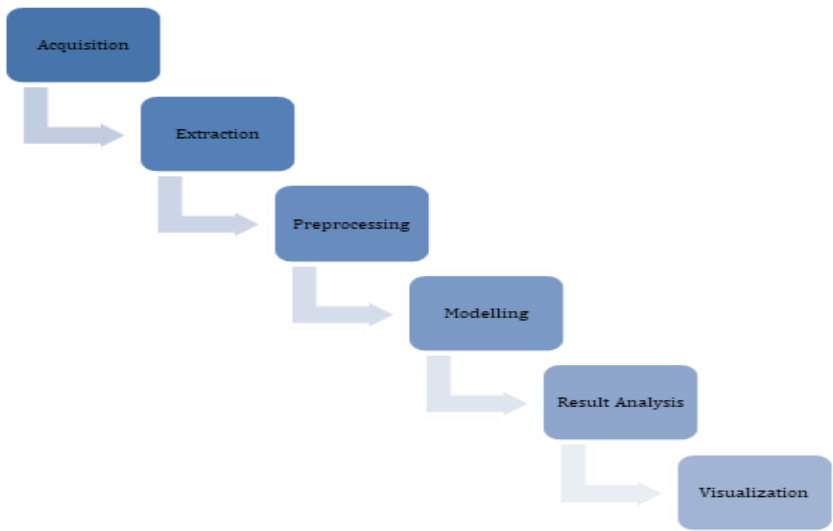

In addition to the above, specific applications have distinctive preprocessing requirements. For instance, removing duplicates when handling author information, is not as straightforward as removing duplicates for numeric or textual entries such as DOI and titles. Author disambiguation is an existing preprocessing challenge with the currently available method, which manifests reasonably accuracy [16]. For most research management applications that focus on paper writing, concept development and references management, preprocessing tasks focus on cleaning text from title, abstract, paper body and/or references [17]. At the end of preprocessing phase, the data is ready to be used as input to the model.

Depending on the application, several modelling techniques are used for the design and development of scholarly applications. Given the textual nature of data, most applications use TF-IDF [18] or PageRank [19] for feature extraction. Common use cases include automatic keyword extraction [20] and topic modelling [21]. With the advancement in machine and deep learning techniques, recent applications have used them for better accuracy.

The coverage of this paper is limited to deep learning based applications and the approaches used to implement them. Therefore, Table 1 presents some recent applications of deep learning for scholarly data and their implementation approaches. All the three applications make use of deep learning techniques, on scholarly data. The implementation details of these applications indicate use of conventional approach. Literature review suggests that Spark [22] has not been used in any capacity for deep learning applications, particularly in the big scholarly data domain.

Table 1. Review of existing deep learning-based applications for big scholarly data

| S. No. | Application | Features | Preprocessing Done | Modelling Technique |
|---|---|---|---|---|
| 1. | PubMender: A system for biomedical venue recommendation [23] | It is a journal recommendation system that works specifically in the biomedical domain. | NLTK for word segmentation | Venue recommendation is considered a multi-label classification problem and CNN is used. |
| 2. | Keyword extraction from scholarly documents using Bi-LSTM-CRF [24] | Solves the key-phrase extraction problem by modelling it as a sequence labelling problem. | Abstract/Key-phrase data pairs are tokenized. | LSTM-CRF, CRF, Bi-LSTM, and LSTM. |
| 3. | Deep Key-phrase Generation [25] | Extracts key-phrases automatically using deep learning techniques | Lowercasing, tokenization and replacing digits with their string versions | RNN and Copy-RNN |

It is found that the following five methods are the most commonly used and are required for textual preprocessing: Removal of punctuation, short words, stopwords, HTML tags and special characters, and others. Finally, the results generated by the model are summarized and presented in the form of textual data or word cloud [26] for better visualization of the results. This paper focuses on preprocessing techniques for deep learning based scholarly data applications. Therefore, this paper will not discuss modelling and vi

## 3. Preprocessing Pipeline for Scholarly Applications (P3SAPP)

As with any deep learning application, the accuracy of the result depends on the amount of data used or training [27]. In other words, the accuracy of the model improves as the size of the dataset increases. Considering the computing requirements of the application, the model can only be appropriately trained with the concerned dataset on a GPU-enabled system.

It is important to note that the increasing size of the dataset for NLP-based applications makes the preprocessing stage of the deep learning model development extremely resource intensive. In the scenario where a cloud-based GPU has been purchased for developing the model, such a proposition can prove expensive as the GPU is underutilized at 0% load during data ingestion and preprocessing stages of model development.

In order to optimize GPU utilization and reduce the total time required for model development, this paper proposes a framework based on an existing preprocessing approach called P3SAPP (Preprocessing Pipeline for Scholarly Applications). The algorithm of the proposed approach is given in Algorithm 1. In order to facilitate comparison of the proposed approach (P3SAPP) with the conventional approach (CA), Algorithm 2, describing the conventional approach, is also provided.

The proposed framework uses a big data technology, Spark [22], for ingesting data and parallelizing the preprocessing stage of model development, reducing the preprocessing time, which in turn reduces the total development time. This reduction has a direct impact on the total time for which a Cloud-based GPU instance shall be required, correspondingly reducing the cost of computing and the overall cost of the project. It is important to mention that this approach improved the data ingestion and preprocessing stages of model development, leaving the last two stages of model training and inference, untouched.

| ALGORITHM 1 P3SAPP | ALGORITHM 2 CA |
|---|---|
| Input: Data files<br>Output: Pandas DataFrame with extracted and cleaned text<br>*BEGIN*<br>1. Initialize a Spark DataFrame, data.<br>2. *FOR* each directory<br>3. *FOR* each file<br>4.    Read file data into a DataFrame<br>5.    Select data to be extracted<br>6.    Perform union between Spark DataFrame 'data' and selected data<br>7. *END FOR*<br>8. *END FOR*<br>9. Remove NULL valued rows<br>10. Remove duplicates<br>11. Define different stages of preprocessing APIs<br>12. Initialize Spark ML Pipeline for preprocessing<br>13. Fit the data on Pipeline<br>14. Transform data using Pipeline<br>15. Convert Spark DataFrame to Pandas DataFrame<br>16. Remove NULL valued rows<br>*END* | Input: Data files<br>Output: Pandas DataFrame with extracted and cleaned text<br>*BEGIN*<br>1. Initialize a Pandas DataFrame, data.<br>2. *FOR* each directory<br>3. *FOR* each file<br>4.    Read file into a DataFrame<br>5.    Select data to be extracted<br>6.    Append the Pandas DataFrame 'data' with selected data<br>7. *END FOR*<br>8. *END FOR*<br>9. Remove NULL valued rows<br>10. Remove duplicates<br>11. FOR all rows in the DataFrame<br>12.    Perform text cleaning<br>13. END FOR<br>14. Remove NULL valued rows<br>*END* |

The overall framework can be broken down into four stages, out of which P3SAPP alters data ingestion and preprocessing stages. Steps 2-8, for both approaches, perform ingestion and time corresponding to their execution is considered as ingestion time. Steps 9-10, for both approaches, perform pre-cleaning and time corresponding to their execution is considered as pre-cleaning time. Steps 11-13, for CA, perform cleaning and time corresponding to their execution is considered as cleaning time for CA. On the other hand, Step 14 performs cleaning for P3SAPP and its execution time corresponds for P3SAPP's cleaning time. Step 14 performs post-cleaning for CA while steps 15-16 perform the same for P3SAPP. The post-cleaning times are correspondingly determined. The total preprocessing time for CA is determined by the execution time of steps 2-14. Correspondingly, execution time for steps 2-16 is used to determine preprocessing time for P3SAPP.

Theoretically, the algorithmic time complexity for conventional approach is $O(n)$ because every element to the concerned column will be accessed and processed. It is noteworthy that n is the total number of elements in the column concerned. On the other hand, for the P3SAPP approach, the time complexity will be $O(\frac{n}{k})$ where k is the number of nodes in the cluster if Spark is operating on cluster mode or the number of cores used to parallelize the job, if Spark is operating on local[*] mode. The details of the proposed approach are described in the following sub-sections.

*3.1 Data ingestion*

Data ingestion is the first stage of model development for any machine learning or deep learning applications. As part of this stage, data is ingested into the system for further processing. Irrespective of the format of base dataset, this approach proposes ingestion of data into a PySpark DataFrame [28].

Since the raw data from a scholarly document is structured in the sense that it can be ingested in the form of rows and columns, Spark SQL has been selected as the base technology for operating with data inside Spark. Spark SQL provides a DataFrame interface, which is capable of operating on different data formats, including JSON, ORC, Parquet and others [29].

Besides, Spark also provides generic data loading and saving methods, in which developers can specify their own working format. This allows the flexibility to work with different formats using the same base technology. The advantage of using a Spark DataFrame is the fact that relational transformation can be performed on data along with the provision to register a DataFrame as a temporary view. Ingestion of data into a Spark DataFrame is more efficient than the ingestion in Panda [30].

*3.2 Data preprocessing*

In view of the requirements of scholarly applications, the ingested data values are typically textual in nature. For most applications that require textual processing, the text needs to be cleaned before it can be sent for further processing. On the basis of literature review, it has been deduced that commonly required text cleaning tasks, including:

      a. Tokenize text
      b. Convert text to lower case
      c. Remove HTML tags, if any

        d. Remove unwanted characters
        e. Remove stopwords
        f. Remove short words

The Spark ML Feature package provides some APIs that are built on top of DataFrames for feature transformation. For text preprocessing, the available APIs includes Tokenizer [31], for tokenizing text and StopWordsRemover [32], for removing stopwords. However, the rest of the APIs are not present and have been implemented in this work.

It is proposed that the APIs must be used to create a Spark ML Pipeline [33] so that Spark can perform the pipelined tasks in a parallel fashion, to reduce the required time. Typically, Spark ML Pipeline consists of transformers and estimators. P3SAPP proposes to use Pipelines for chaining multiple transformer APIs to specify a preprocessing workflow. On the basis of the preprocessing requirements, different transformer APIs can be chosen and chained in the pipeline for faster preprocessing.

Finally, the resulting Spark DataFrame is transformed into a Pandas DataFrame, which can be fed to the model training sub-system. This sub-step is in line with the black box model of development. The proposed approach does the same as the conventional approach, which takes raw data as the input, then generates Pandas frame as the output for the subsequent model development.

*3.3 Model training and inference*

The model shall be developed on the basis of the required application and trained using the generated Pandas DataFrame. The trained model can then be inferred to deduce the required results.

**4. Methodology**

This paper presents the implementation of the Spark ML Feature [33] APIs mentioned in Section 4.1. In order to create a preprocessing workflow, Spark Pipeline [33] was used. P3SAPP was tested using a case study that generates titles for scholarly works on the basis of the provided abstract. The details of the case study are provided in Section 4.2.

*4.1 Spark ML feature APIs*

As mentioned previously, Spark ML Feature [33] package already has APIs developed for tokenizing text and removing stopwords. The class has predefined methods for implementation of initialization and transformation functions. Such methods are used for implementation of fours APIs used in this work, which are mentioned below. The code for the implemented APIs is available at GitHub [48].

4.1.1 ConvertToLower

This API performs case conversion of the all the row entries for the column provided as input. The case of all the alphabets in the entries is changed to lowercase. This API is essential in view of the fact that most NLP tasks require matching, similarity identification or manipulation based on identification of alphabets, words or strings. The use of such an API reduces programming effort by bringing all the values on the same level of casing.

4.1.2 RemoveHTMLTags

In view of the fact that the primary source of all scholarly data is the web and in most scenarios, it is required to crawl the web and retrieve data, textual data is typically retrieved as HTML content with tags. Although, this may or may not be true for all entries, it can be taken as a mandatory text-cleaning step before any analytical task can be performed.

4.1.3 RemoveUnwantedCharacters

Once all the text is in lowercase and devoid of all tags, string-based manipulation can be performed. Common cleaning tasks require removal of the following characters or textual elements.
- Punctuation
- Text between parentheses
- Apostrophes
- Digits and any special characters
- Performs contraction mapping

This API performs textual cleaning by removing all the above-mentioned textual elements and outputs strings that have relevant words and phrases for advanced processing.

4.1.4 RemoveShortWords

Some words like abbreviations or conjunctions that are not typically removed using other APIs can be identified and removed on the basis of their word length. As part of this API, the user is expected to provide another input named threshold, which determines the maximum number of characters that a word should have for it to be considered for removal. Therefore, this API removes all words that are equal to or less than the threshold value in length.

*4.2  Case study: Deep learning – based title generation for scholarly articles*

In order to test the feasibility and degree of time and cost reduction of using the proposed approach, a case that cleans texts in different capacities was chosen. Title or summary generation from abstracts for scholarly articles is an application that can be used in many different ways. Some of the common objectives include paper review management system, which can generate summary of received articles to facilitate editorial decision on a manuscript and research paper writing application that can automatically suggest appropriate titles for a scholarly article on the basis of provided abstract.

4.2.1 Data Ingestion

The data, which is available in the form of JSON files, is ingested into Spark DataFrame using API provided for the same. Only data corresponding to titles and abstracted is ingested.

4.2.2 Preprocessing Pipeline

The preprocessing stage is divided into three sub-stages namely, pre-cleaning, cleaning and post-cleaning. For the conventional approach, the three stages perform the following functions –
- The pre-cleaning stage removes nulls and duplicates.
- The cleaning stage performs different set of operations on titles and abstracts. For abstracts, text is converted to lowercase and HTML Tags, Unwanted Characters, Stopwords and short words are removed. On the other hand, for abstracts, text is converted to lowercase     and HTML Tags, Unwanted Characters and short words are removed. The implemented APIs,

ConvertToLower RemoveHTMLTags, RemoveUnwantedCharacters and RemoveShortWords were used. Moreover, although, StopWordsRemover is a generic API available for stopwords removal, the case study - specific implementation for the same was also done.
- The post cleaning stage again checks for any nulls and removes them.

At the end of the post-cleaning stage, the Pandas dataframe is ready to be imported into the Model Training module. The proposed approach performs the same set of steps for the three different stages of preprocessing. However, all the transformational operations are performed on Spark DataFrame and in the post-cleaning stage, the Spark DataFrame is converted to a Pandas DataFrame. The preprocessing workflows required for abstracts and titles are different and shown in Fig. 2 and Fig. 3. Since, the abstract will be used as feature for training the model, it must be completely clean. Therefore, the cleaning tasks performed for abstracts include –
- Convert all the text to lowercase
- Remove all HTML Tags if any.
- Remove all unwanted characters.
- Remove stopwords.
- Remove short words.

On the other hand, title is the target for the model and thus, the cleaning tasks required include –
- Convert all the text to lowercase
- Remove all HTML Tags if any.
- Remove all unwanted characters.

For the purpose of case study implementation, the threshold value for short words removal is fixed at threshold = 1. The respective APIs are called to define a Spark ML pipelines. The pipelines are fitted to data and the input dataframe is transformed using this pipeline.

Fig. 2. Spark ML-based preprocessing pipeline for cleaning abstracts

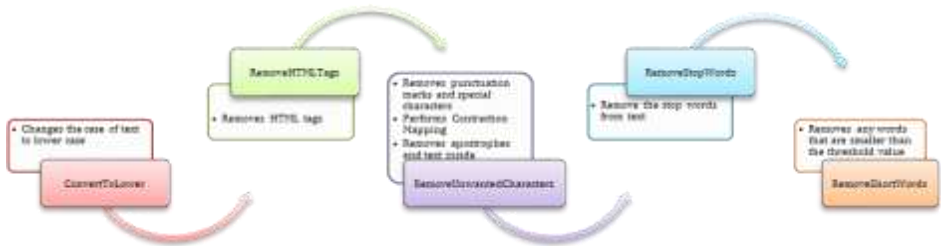

4.2.3 Model training and inference

The case study will showcase the implementation of text summarization for scholarly articles. There are two types of text summarization methods namely, abstractive text summarization [36] and extractive text summarization [37]. Extractive text summarization identifies and extracts sentences, phrases and words from the original text, while abstractive text generates new sentences that summarize the original text. The problem of title generation from abstract of scholarly article requires abstractive text summarization.

Fig. 3. Spark ML-based preprocessing pipeline for cleaning titles

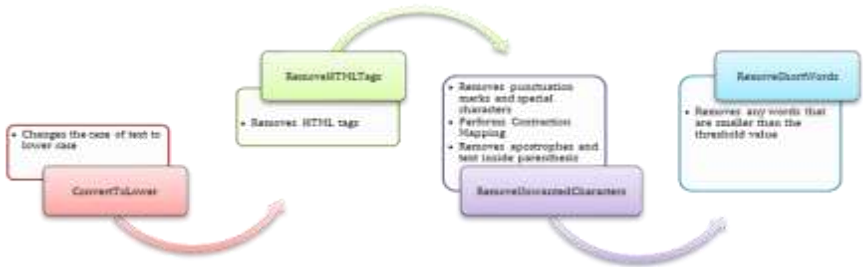

Fig. 4. Training phase: LSTM encoder architecture [42]

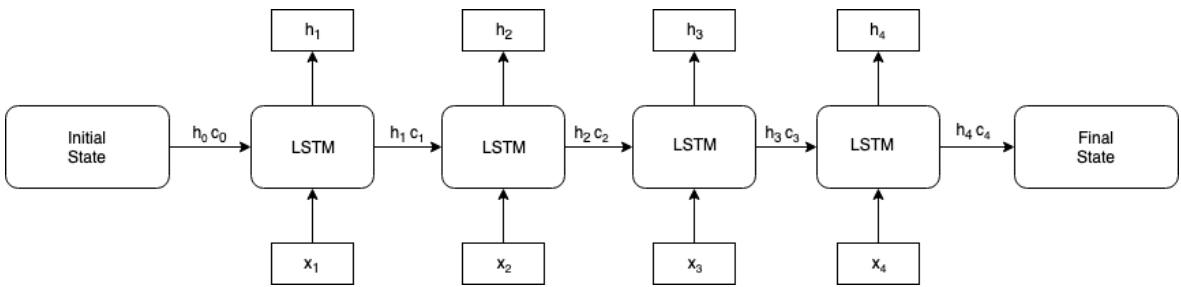

Fig. 5. Training phase: LSTM decoder architecture [42]

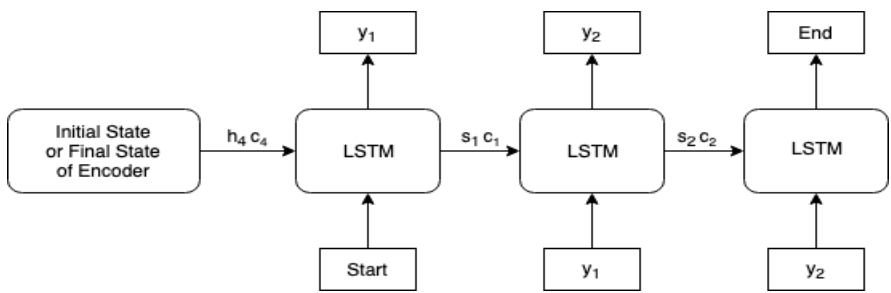

  Text is sequential information and text summarization requires seq2seq modelling [38] where the input sequence is a long text while the output sequence is its summary or short text. Therefore, generating title from abstract of a scholarly article is a many-to-many seq2seq problem. The seq2seq model is composed of two components namely encoder and decoder. The encoder and decoder are generally implemented using variants of Recurrent Neural Networks (RNN) [39] such as Long Short Term Memory (LSTM) [40] or Gated Recurrent Neural Network (GRU) [41]. The reason for this assertion is that RNNs are better capable of managing the issue related to vanishing gradient. As a result, they can capture long term dependencies more efficiently.

The setting up of the encoder and decoder is divided into two phases namely training phase and inference phase. The details of implementation are as follows:

- Training

  In the training phase, the encoder and decoder setup is performed. The model is trained to make a prediction of the target sequence offset per time-step. Therefore, at each time-step, the encoder LSTM processes the input sequence to feed one word into the encoder. The encoder's job is to comprehend and learn the input sequence's contextual information. The encoder architecture is illustrated in Fig. 4. It is important to note that $h_i$ and $c_i$ are the hidden and cell states respectively. Since, encoder and decoder are different stages, the hidden and cell states are fed to the decoder for initialization.

  It is the decoder's job to read target sequence and make predictions on the basis of sequence offset per time-step. Therefore, every next word is predicted using the previous word. The flow chart for decoder architecture is illustrated in Fig. 5. Since the target sequence's first word is unknown, the first word passed to the decoder is <start> token. Moreover, the <end> token marks the end of sentence.

  In order to build a model, a 3-layer stacked LSTM is used for encoder. Using a stacked LSTM ensures better sequence representation. The model is instructed to stop early when the validation loss begins to increase. This is performed to optimise the number of epochs executed for model building.

- Inference

  The encoder and decoder of LSTM are setup for the inference stage. Fig. 6 illustrates the model inference architecture. The steps for model inference are provided in Algorithm 4.

| **ALGORITHM 3** |
|---|
| **Model Inference** |
| Input: <start> token |
| Output: Generated text string |
| *BEGIN* |
| 1. The entire input sequence is encoded. The generated internal states are fed to the decoder for initialization. |
| 2. The <start> token is given as input. |
| 3. The decoder is run for one time-step. |
| 4. The next is determined with probability of occurrence. The word with the highest probability is chosen. |
| 5. The generated word is passed as input to the decoder for next time-step. The internal states are also updated according to the time-step. |
| 6. Steps 3-5 are repeated until maximum limit of word generation is reached or <end> is generated. |

END

There are certain limitations of the training architecture. The job of an encoder is to convert the complete input sequence into a vector of fixed length. This approach works well for short sequences. However, when dealing with long sequences, the model may suffer from inability to memorize the input sequence into a fixed length vector. In order to solve this problem, attention mechanism [43], which modifies the approach in the sense that the model is now attentive to important sequences in the input instead of memorizing the whole input sequence.

Fig. 6. Inference phase: Architecture [42]

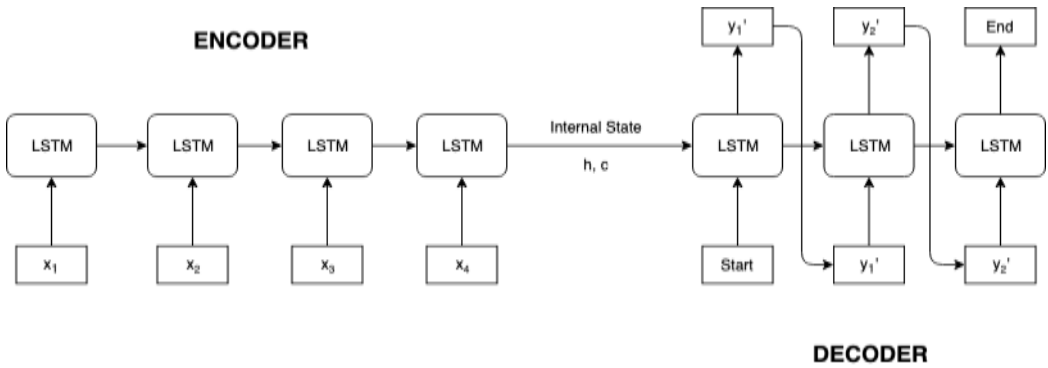

As seen in Fig. 4 and Fig. 5, the encoder generates a hidden state $h_j$ for every j time-step and decoder generates the hidden state $s_i$ for every i time-step. The alignment of the source word (alignment score or $e_{ij}$) with the target word is calculated using the score function, which is given by equation (1):

$$e_{ij} = score(s_i, h_j) \tag{1}$$

There are many types of score function like dot product, additive and general. Once the alignment score is calculated, the softmax function is used for normalizing the scores and getting attention weights. Equation (2) describes the mathematical computation for attention weights ($a_{ij}$).

$$a_{ij} = e^{e_{ij}} \sum_{k=1}^{T_x} e^{e_{ik}} \tag{2}$$

A linear sum of products is computed with attention weights and encoder's hidden states to determine attended context vector ($C_i$), which is given by equation (3).

$$C_i = \sum_{j=1}^{T_x} a_{ij} h_j \tag{3}$$

The attended hidden vector ($S_i$) is computed by concatenating the attended context vector and decoder's target hidden state for time-step $i$ and is given by equation (4).

$$S_i = concatenate([s_i; C_i]) \tag{4}$$

The attended hidden vector $S_i$ is given to the dense layer for computation of $y_i$. Therefore, $y_i$ is given by equation (5).

$$y_i = dense(S_i) \tag{5}$$

The implementation of the text summarization model is inspired by Pai's Keras implementation [42] for text summarization. Since Keras does not have an inbuilt attention mechanism, Ganegedara's implementation [44] of Bahdanau Attention mechanism [43] has been used.

5. **Results and evaluation**

In order to implement a deep learning model for text summarization, a dataset with titles and abstracts was chosen. For this paper, the CORE dataset [49] with the schema shown below was selected because of its easy availability. The full dataset is a zipped file of 330 GB data size. The unzipped version expands to 1.44 TB. It includes 123M metadata items with 85.6M items containing abstracts. The structure of data for each record is as follows:

```
{
  "doi": str|None,
  "coreId": str|None,
  "oai": str|None
  "identifiers": [str],
  "title": str|None,
  "authors": [str],
  "enrichments": {
    "references": [str],
    "documentType": {
      "type": str|None,
      "confidence": str|None
    }
  },
  "contributors": [str],
  "datePublished": str|None,
  "abstract": str|None,
  "downloadUrl": str|None,
  "fullTextIdentifier": str|None,
```

```
  "pdfHashValue": str|None,
  "publisher": str|None,
  "rawRecordXml": str|None
  "journals":[str],
  "language": str|None,
  "relations": [str],
  "year": int|None,
  "topics": [str],
  "subjects": [str],
  "fullText": str|None
}
```

The complete dataset includes 2085 JSON files of variable size. For the purpose of this paper, five subsets with the following files, was created. The sizes of the datasets used for the five use cases are 4.18 GB, 8.54 GB, 13.34 GB, 18.23 GB and 23.58 GB. The files are selected in such a manner that datasets are composed of different number of files, with each file of variable size, ranging from sizes of the order of KB to GB. Moreover, an incremental approach is used for increasing the dataset size because a completely changed dataset may induce a changed behaviour from the system.

The development and testing environment makes use of a GPU of the following configuration: Tesla K80 – 12 GB Memory and 61 GB RAM – 100 GB SSD. The CPU used was Intel Xeon with 2 cores; 8 GB Memory and 200 GB SSD. FloydHub provisioned the GPU and CPU from a cloud-based environment. Spark version v2.4.4 in local [*] mode was used for generation of results. The testing and evaluation of the P3SAPP intends to capture the variations in development time and accuracy for proposed as well as conventional approaches. Finally, the obtained results are used to estimate the impact of P3SAPP on the cost of the project.

*5.1 Execution time*

The total time required for execution of a deep learning application can be defined as follows:

$$T = t_i + t_{pp} + (n * t_{mt}) + t_{mi} \tag{6}$$

The variables used in equation (6) are as follows:

$T$ = Total execution time
$t_i$ = Data ingestion time
$t_{pp}$ = Preprocessing time
$n$ = Number of epochs
$t_{mt}$ = Model training time
$t_{mi}$ = Model inference time

In all experiments, value of $t_{mi}$ for generating a single summary was approximately the same, with the following value:

$$t_{mi} \sim 2\ seconds$$

The value of $t_{mi}$ is negligible in comparison to $t_i$, $t_{pp}$ and $t_{mt}$. Therefore, the value of $t_{mi}$ is ignored for total time computation and cost analysis. Besides this, cumulative time ($t_c$) is defined as follows:

$$t_c = t_i + t_{pp} \qquad (7)$$

Thus, the revised equation is follows:

$$T = t_c + (n * t_{mt}) \qquad (8)$$

The proposed approach reduces cumulative time ($t_c$); the results for which are provided in the sections given below.

5.1.1 Ingestion Time

Ingestion time is defined as the time to ingest data from multiple JSON files into a Spark DataFrame. For this case study, the system ingests values for titles and abstracts from JSON files into a Spark DataFrame. The values of ingestion time determined in the performed experiments are given in Table 2.

Fig. 7 illustrates variations in ingestion time with rise in dataset size. While the conventional approach shows staggering growth with ingestion time shooting up for higher dataset sizes, P3SAPP manifests a slower increase in ingestion time with increase in dataset size. Moreover, ingestion time is reduced by more than 99% for datasets larger than 5 GB.

Table 2. Comparison of Ingestion Time for CA and P3SAPP

| Dataset ID | Dataset Size (GB) | Ingestion Time (sec) | | |
|---|---|---|---|---|
| | | CA | P3SAPP | Reduction (%) |
| 1 | 4.18 | 433.631 | 13.076 | **96.984** |
| 2 | 8.54 | 3542.393 | 26.253 | **99.259** |
| 3 | 13.34 | 8701.101 | 79.843 | **99.082** |
| 4 | 18.23 | 17139.434 | 93.637 | **99.454** |
| 5 | 23.58 | 32698.916 | 104.055 | **99.682** |

Fig. 7. Analysis of Ingestion Time

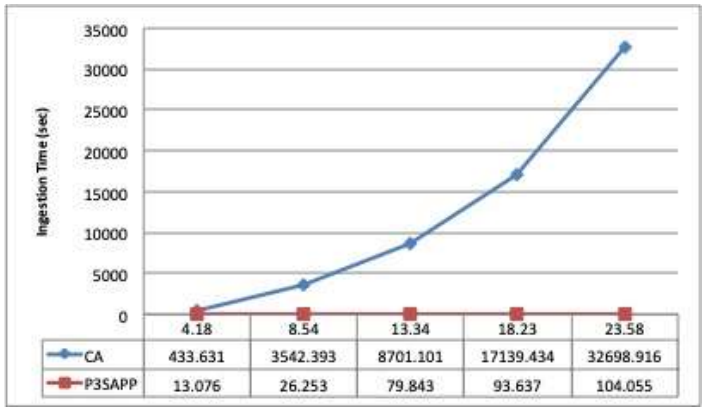

*16*

Table 3. Comparison of Preprocessing Time for Conventional Approach and P3SAPP

| Dataset ID | Dataset Size (GB) | Pre-Cleaning (sec) | | Cleaning (sec) | | Post-Cleaning (sec) | | Total Preprocessing Time (sec) | | |
|---|---|---|---|---|---|---|---|---|---|---|
| | | CA | P3SAPP | CA | P3SAPP | CA | P3SAPP | CA | P3SAPP | Reduction (%) |
| 1 | 4.18 | 0.165 | 0.009 | 154.394 | 0.161 | 0.118 | 89.31 | 154.679 | 89.485 | **42.148%** |
| 2 | 8.54 | 0.273 | 0.008 | 232.223 | 0.154 | 0.247 | 140.442 | 232.745 | 140.609 | **39.589%** |
| 3 | 13.34 | 0.528 | 0.008 | 457.768 | 0.172 | 0.452 | 262.307 | 458.94 | 262.492 | **42.800%** |
| 4 | 18.23 | 0.811 | 0.017 | 628.464 | 0.206 | 0.635 | 351.62 | 629.913 | 351.848 | **44.143%** |
| 5 | 23.58 | 1.067 | 0.017 | 862.453 | 0.252 | 0.887 | 477.51 | 864.409 | 477.784 | **44.727%** |

Fig. 8. Analysis of Preprocessing Time

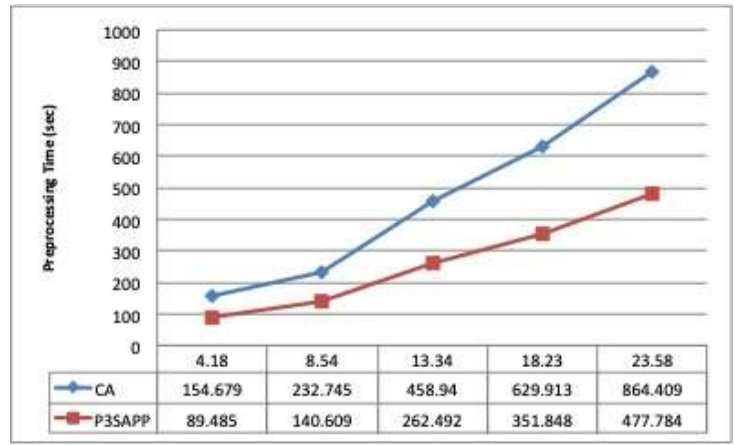

### 5.1.2 Preprocessing Time

Preprocessing time is the total time required by the system to clean ingested data. The total preprocessing time is computed using the following equation (4). The values of pre-cleaning, cleaning, post-cleaning and total preprocessing time determined in the performed experiments are given in Table 3. Fig. 8 illustrates the trends for preprocessing times obtained for conventional and proposed approaches.

The rise in preprocessing time for conventional approach is steeper than the same obtained for the proposed approach, exhibiting an average reduction of approximately 40%. It is important to note that cleaning stage takes a chunk of the time for the conventional approach. On the other hand, conversion of Spark DataFrame to Pandas DataFrame in the post-cleaning stage consumes most of the total preprocessing time for the proposed approach. The results determined in the performed experiments are given in Table 3. Fig. 6 illustrates variations in preprocessing time with rise in dataset size.

$$t_{pp} = t_{prc} + t_c + t_{poc} \quad t_{pp} = t_{prc} + t_c + t_{poc} \quad \text{-----------------------(9)}$$

### 5.1.3 Cumulative Time

Cumulative time is sum of ingestion and preprocessing time and is calculated using equation (2). The trend for cumulative time obtained using conventional approach exhibits staggering rise while the proposed approach manifests a very slow rise. The reduction in cumulative time is increasing with increase in data size, making this approach more beneficial for larger datasets. The values of cumulative time determined in the performed experiments are given in Table 4. Fig. 9 illustrates variations in cumulative time with rise in dataset size.

Table 4 Comparison of Cumulative Time for Conventional Approach and P3SAPP

| Dataset ID | Dataset Size (GB) | Total Time (sec) | | |
|---|---|---|---|---|
| | | CA | P3SAPP | Reduction (%) |
| 1 | 4.18 | 588.31 | 102.561 | **82.567** |
| 2 | 8.54 | 3775.138 | 166.862 | **95.58** |
| 3 | 13.34 | 9160.041 | 342.335 | **96.263** |
| 4 | 18.23 | 17769.347 | 445.485 | **97.493** |
| 5 | 23.58 | 33563.325 | 581.839 | **98.266** |

Fig. 9. Analysis of Cumulative Time

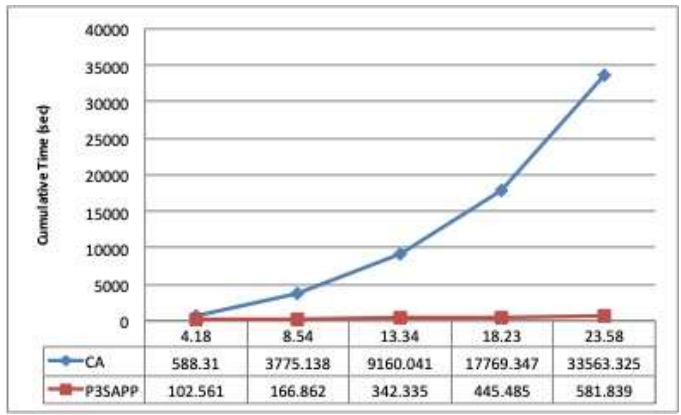

## 5.2 Accuracy

The accuracy for the proposed approach, P3SAPP, is determined by the percentage of matching records in the Panda DataFrames generated for conventional and proposed approaches. The extracted records in the form a Pandas DataFrame for both the approaches were compared to determine the matching records and consequently, the percentage of matching records. The results obtained for accuracy are provided in Table 5 and Table 6. The average accuracy for titles was determined to be 96.595%. On the other hand, the average accuracy for abstracts was found to be 97.929%.

Table 5. Matching Records for Extracted Titles

| Dataset ID | Conventional Approach | Proposed Approach | Matching Records | Percentage |
|---|---|---|---|---|
| 1 | 88709 | 88709 | 86935 | 98.00020291 |
| 2 | 132683 | 132683 | 128924 | 97.16693171 |
| 3 | 256362 | 256362 | 248950 | 97.10877587 |
| 4 | 345169 | 345169 | 334881 | 97.01943106 |
| 5 | 480712 | 480713 | 450333 | 93.68041572 |

Table 6. Matching Records for Extracted Abstracts

| Dataset ID | Conventional Approach | Proposed Approach | Matching Records | Percentage |
|---|---|---|---|---|
| 1 | 88709 | 88709 | 88282 | **99.518%** |
| 2 | 132683 | 132683 | 129179 | **97.359%** |
| 3 | 256362 | 256362 | 251572 | **98.132%** |
| 4 | 345169 | 345169 | 339541 | **98.369%** |
| 5 | 480712 | 480713 | 462766 | **96.267%** |

## 5.3 Cost Benefit

Cloud-based services like AWS [45], GCP [46] and FloydHub [47], provision Platform-as-a-Service (PaaS), on hourly expenditure. Therefore, the total cost can be estimated on the basis of the number of hours a job will take to complete. The total time for conventional and proposed approach can be computed using equation (8). For cost benefit evaluation, the number of epochs is fixed as 10, 25 and 50.
From equation (8):

$$T = t_c + (n * t_{mt})$$

The cost benefit is determined by converting total time in hours and multiplying the value with hourly cost. The formula for cost evaluation is given by,

$$C = x * TC = x * T \quad \text{-----------(10)}$$

In equation (10), $C$ is the total cost of execution and $x$ is hourly cost. Using equation (10), cost benefit can be given by,

$$CB = \frac{x * (T_{ca} - T_{pa})}{x * T_{ca}} * 100$$
$$CB = \frac{(T_{ca} - T_{pa})}{T_{ca}} * 100 \qquad \qquad \text{-----------------------------(11)}$$

In equation (6), $CB$ is Cost Benefit, $T_{ca}$ is total time taken for conventional approach and $T_{pa}$ is total time taken for proposed approach. The results of the computation performed for determination of $T$ and $CB$ are provided in Table 7. The results indicate a rise in cost benefit with increase in dataset size. However, as the number of epochs increase, the corresponding cost benefit is lowered, as is evident from Fig. 11.

6. **Discussion**

As mentioned previously, the theoretical time complexity of preprocessing phase of the algorithm that is escalated to Spark is $O(n)$ for the conventional approach and $O(\frac{n}{k})$ for P3SAPP, which means that the execution time for the proposed approach is expected to scale down by a factor k, which depends on the Spark mode and number of nodes on the cluster. The experimental results validate this assertion. Fig. 10 shows the graphs for preprocessing results of CA and P3SAPP respectively. It can be seen that both the graphs exhibit a linear relationship between preprocessing time and dataset size. Moreover, the equations of the graphs indicate that for every unit increase in dataset size, the preprocessing time increases 37.589 times for CA while the same for P3SAPP occurs by a factor of 20.426. The values of slope are expected to change depending on the Spark configuration used.

Fig. 10. Trend-line Graphs of CA and P3SAPP for Preprocessing Results

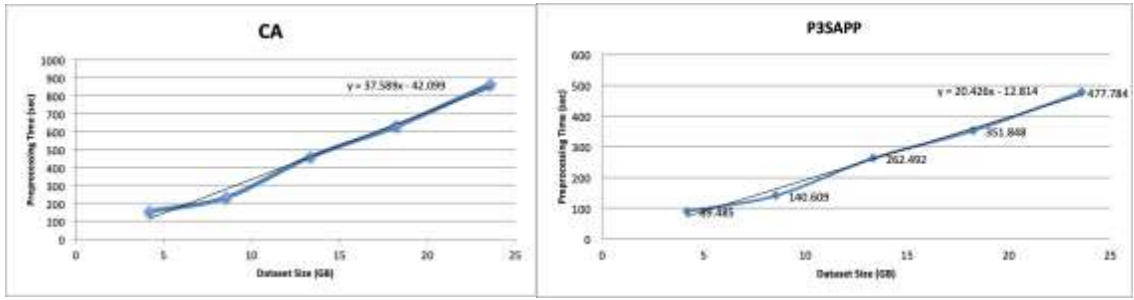

The analysis of results covers three significant aspects namely execution time, accuracy and cost-benefit. The proposed approach offers major time benefits, providing more than 90% reduction for datasets larger than 5 GB with major savings in ingestion time and a consistent saving of around 40% in preprocessing time. Fig. 12 provides a summarization of percentage reduction in ingestion, preprocessing and cumulative time, with respect to the five datasets considered for experimentation.

Table 7. Cost-Benefit Analysis

| Dataset ID | Cumulative Time (secs) | | MTT per epoch (secs) | Total Time for 10 epochs (hrs) | | | Total Time for 25 epochs (hrs) | | | Total Time for 50 epochs (hrs) | | |
|---|---|---|---|---|---|---|---|---|---|---|---|---|
| | CA | P3SAPP | | CA | P3SAPP | Cost Benefit | CA | P3SAPP | Cost Benefit | CA | P3SAPP | Cost Benefit |
| 1 | 588.31 | 102.561 | 1132 | 3.310 | 3.173 | **4.079%** | 8.024 | 7.890 | **1.681%** | 15.886 | 15.751 | **0.849%** |
| 2 | 3775.138 | 166.862 | 1698 | 5.765 | 4.763 | **17.385%** | 12.840 | 11.838 | **7.805%** | 24.632 | 23.630 | **4.069%** |
| 3 | 9160.041 | 342.335 | 3166 | 11.339 | 8.889 | **21.601%** | 24.530 | 22.081 | **9.985%** | 46.517 | 44.067 | **5.265%** |
| 4 | 17769.347 | 445.485 | 4070 | 16.241 | 11.429 | **29.629%** | 33.120 | 28.388 | **14.495%** | 61.464 | 56.651 | **7.829%** |
| 5 | 33563.325 | 581.839 | 4170 | 20.906 | 11.745 | **43.821%** | 39.440 | 29.120 | **26.166%** | 67.240 | 58.078 | **13.625%** |

Fig. 11. Cost Benefit Analysis

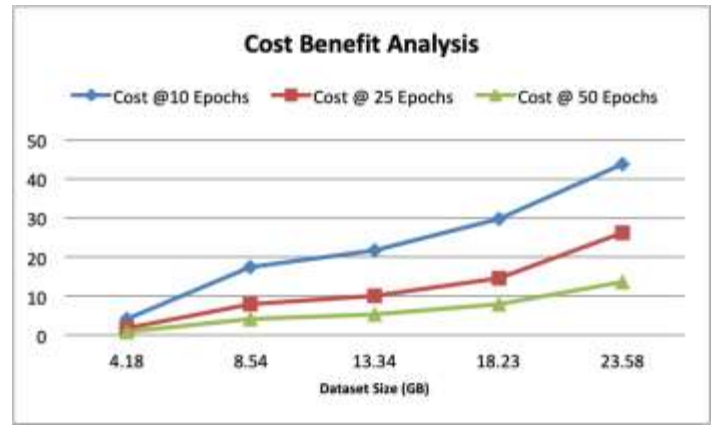

Fig. 12. Development time - Summary of results

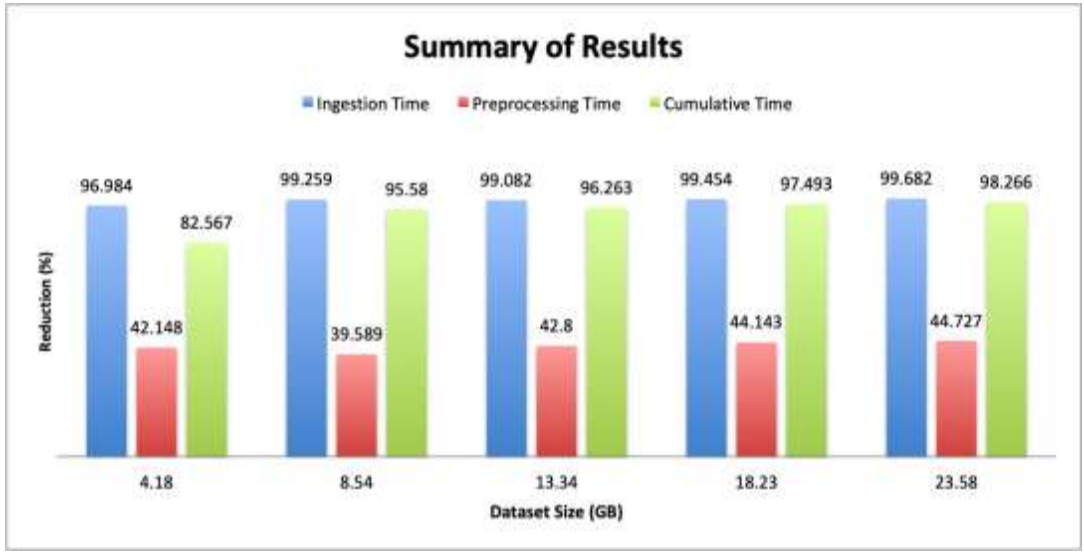

In view of the fact that Model Training Time (MTT) is the most time-consuming activity, a comparison between time saving and model training time per epoch is necessary. The summary of findings in this regard is provided in Table 8. Evidently, the ratio of time saving and MTT/epoch increases exponentially with rise in size of dataset (as shown in Fig. 13). Moreover, for Dataset ID = 5, this value is as high as 7.9, which means the time savings provided by the proposed approach is equal to the time taken by 7.9 epochs. The significance of this value can translate into major time and cost savings for projects that work with larger datasets.

Table 8. Reduction in Preprocessing Time in terms of Model Training Time (MTT) per epoch

| Dataset ID | Dataset Size (GB) | Number of Training Records | Number of Validation Records | MTT per epoch (sec) | Time Saving (sec) | Ratio of Time Saving and MTT per epoch |
|---|---|---|---|---|---|---|
| 1 | 4.18 | 70505 | 7834 | 1132 | 485.749 | 0.429 |
| 2 | 8.54 | 104368 | 11597 | 1698 | 3608.296 | 2.125 |
| 3 | 13.34 | 200908 | 22327 | 3166 | 9160.041 | 2.893 |
| 4 | 18.23 | 270514 | 30023 | 4070 | 17323.862 | 4.256 |
| 5 | 23.58 | 383002 | 42536 | 4170 | 32981.486 | 7.909 |

Fig. 13. Ratio of Time Saving and MTT/Epoch

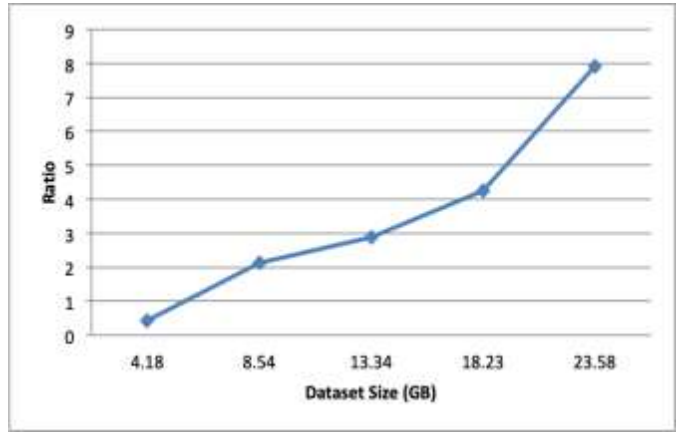

Results indicate that cost benefit is expected to rise with increase in dataset size for a fixed number of epochs. Although, the proposed approach records high accuracy in terms of matching records produced by the two approaches, it is noteworthy that accuracy reduces for larger datasets, but remains more than 93%. The reason for non-matches between records from the two DataFrames can be attributed to the difference in ingestion methods. Reduction in this parameter and the impact in variations in matching records on the generated model shall be studied in the future.

Another important point to note is that the proposed approach has been implemented and tested with Spark on local [*] mode, which means that Spark is running locally and the different worker threads are working on the different logical cores on the machine. Spark can be moved to a full-fledged cluster to get enhanced results. Although, this work does not present any such results, it can be done in the future.

Moreover, escalating the deep learning model to Spark can also be explored in the future to reduce development time and costs, further. As it can be seen in Table 3, the variations in the total preprocessing time arises due the post cleaning time in the conversion of Spark Dataframe to Pandas Dataframe. Escalation of model to Spark shall remove this aspect of the proposed approach. Therefore, for a given configuration of Spark, the preprocessing time, in such a scenario, will be constant.

## 7. Conclusion

This work proposes a framework that modifies the preprocessing stages of deep learning model development. The proposed approach backs the use of Spark for ingesting data and parallelizing the different preprocessing tasks required for scholarly data applications. Preprocessing for deep learning model development, particularly for scholarly applications, is highly resource intensive. As a result, for application development that uses Cloud provisioned platforms and infrastructure, most of the time is wasted, as GPU remains underutilized during this time. The GPU is utilized only when Tensorflow is used. Therefore, reducing the preprocessing time reduces the time of underutilization and overall cost of the project.

The proposed approach provides more than 90% reduction in total preprocessing time, which includes ingestion and preprocessing time, for datasets larger than 5 GB. Besides this, the cost saving are dependent on the number of epochs and size of datasets. Cost savings are highest for lesser epochs and large datasets. It is important to note that both cost saving and reductions in cumulative time increase with increase in dataset size, making this approach highly relevant for big datasets. This shall also improve the accuracy of the developed deep learning model, as larger datasets are known to exhibit better results.

The accuracy the approach in terms of matching records obtained when compared to conventional approach is more than 90% for datasets larger than 5 GB. The cause of mismatches is rooted in differences in ingestion. Further investigations and efforts to improve results for this aspect of the approach shall be attempted in the future. As part of this work, four APIs were implemented for enhancing the Spark ML Feature Class. More APIs can be identified and implemented in the future. The proposed model has used Spark on local [*] model, which parallelizes different threads on different logical cores. Higher levels of parallelization can be investigated in future work. Moreover, escalation of the deep learning model to Spark will also be explored in the future.

**Acknowledgements**

This work was supported by a grant from "Young Faculty Research Fellowship" under Visvesvaraya PhD Scheme for Electronics and IT, Department of Electronics & Information Technology (DeitY), Ministry of Communications & IT, Government of India.